
%
%
%
%
\input harvmac.tex

\def\frac#1#2{{\textstyle{#1\over #2}}}

\def\b#1{\kern-0.25pt\vbox{\hrule height 0.2pt\hbox{\vrule
width 0.2pt \kern2pt\vbox{\kern2pt \hbox{#1}\kern2pt}\kern2pt\vrule
width 0.2pt}\hrule height 0.2pt}}

\def\STrow#1{\hbox{#1}\kern-1.35pt}

\def\tri#1#2#3#4#5#6#7#8#9{\matrix{#4\cr
	#3\quad#5\cr #2~\qquad #6\cr #1\quad #9\quad#8\quad#7\cr}}

\def\tria#1#2#3#4#5#6#7#8{\matrix{\quad\cdot #1 \qquad #5 \cdot \quad \cr
#4 \cdot\quad\cdot #2 \qquad #6 \cdot\quad\cdot #8 \cr
\quad\cdot #3 \qquad #7 \cdot\quad \cr}}


\def\trianb#1#2#3#4#5#6#7#8{\matrix{\quad&\quad&\quad&\cdot#1&\quad&
\quad&#5\cdot\quad&\quad&\quad&\quad \cr
 \cr
#4\cdot\quad&\quad&\quad&~~\cdot#2\quad&\quad&\quad&#6\cdot\quad&\quad&
\quad&\cdot#8 \cr
\cr
\quad&\quad&\quad&~~\cdot#3\quad&\quad&\quad&#7\cdot\quad&\quad&\quad&\quad}}

\def\text#1{\quad\hbox{#1}\quad}

\def\la{\lambda}

\def\su{\widehat{su}}
\def\Nb{N_{\lambda\mu\nu}}
\def\LLL{\lambda_{1}+\mu_{1}+\nu_{1}}
\def\LLM{\lambda_{2}+\mu_{2}+\nu_{2}}
\def\kmin{k_{0}^{min}}
\def\kmax{k_{0}^{max}}
\def\E1{E_{1}}
\def\max{{\rm max}}
\def\min{{\rm min}}
\overfullrule=0pt

\newcount\eqnum
\eqnum=0
\def\eq{\eqno(\secsym\the\meqno)\global\advance\meqno by1}
\def\eqlabel#1{{\xdef#1{\secsym\the\meqno}}\eq }

\newwrite\refs
\def\startreferences{
 \immediate\openout\refs=references
 \immediate\write\refs{\baselineskip=14pt \parindent=16pt \parskip=2pt}
}
\startreferences

\refno=0
\def\aref#1{\global\advance\refno by1
 \immediate\write\refs{\noexpand\item{\the\refno.}#1\hfil\par}}
\def\ref#1{\aref{#1}\the\refno}
\def\refname#1{\xdef#1{\the\refno}}

\def\pl#1#2#3{Phys. Lett. {\bf B#1#2#3}}
\def\np#1#2#3{Nucl. Phys. {\bf B#1#2#3}}

\newcount\exno
\exno=0
\def\Ex{\global\advance\exno by1{\noindent\sl Example \the\exno:

\nobreak\par\nobreak}}

\parskip=6pt
\Title{\vbox{\baselineskip12pt
\hbox{LAVAL-PHY-22/92}\hbox{LETH-PHY-5/92}}}
{\vbox {\centerline{$\widehat{su}(3)_k$ fusion coefficients}
}}
\centerline{L. B\'egin$^\natural$\foot{Work supported by NSERC
(Canada).}, P. Mathieu$^\natural$\foot{Work supported
by NSERC (Canada) and FCAR (Qu\'ebec).} and M.A. Walton$^{\sharp 1}$}
\vskip.2in
\smallskip\centerline{$^\natural$ \it D\'epartement de Physique,
Universit\'e Laval, Qu\'ebec, Canada G1K 7P4}
\smallskip\centerline{$^\sharp$ \it Physics Department, University of
Lethbridge,
Lethbridge (Alberta) Canada T1K 3M4}
\vskip .2in
\centerline{\bf Abstract}
\bigskip
\noindent
A closed and explicit formula for all $\su{(3)}_k$ fusion coefficients
is presented which, in the limit $k \rightarrow \infty$, turns into a simple
and compact expression for the $su(3)$ tensor product coefficients.  The
derivation is based on a new diagrammatic method which gives directly both
tensor product and fusion coefficients.
\Date{5/92\ \ (hepth@xxx/9206032)}


Fusion rules in WZNW models can be calculated rather efficiently from tensor
product coefficients by means of the affine Weyl group [\ref{M.A. Walton,
\np340
(1990) 777; \pl241 (1990) 365;V. Ka\v c, {\it Infinite dimensional Lie
algebras}, 3rd ed.  Cambridge University Press},\ref{ F. Goodman and H. Wenzl,
Adv. Math. {\bf 82} (1990) 244.}\refname\GOW,\refname\KW\ref{M. Spiegelglas,
\pl245 (1990) 169; P. Furlan, A. Ganchev and V.B. Petkova, \np343 (1990) 205;
J.
Fuchs and P. van Driel, \np346 (1990) 632.}].  However, closed and explicit
formulae for  fusion coefficients have been obtained previously only in the
$su(2)$ case.   Here we display an explicit expression for the $\su{(3)}_k$
fusion
 coefficients.  This result is the culmination of a number of recent works
in which various combinatorial methods have been devised for the computation of
$\su {(3)}_k$ fusion rules [\ref{C.J. Cummins, J.Phys. A:Math. Gen. \pl24
(1991) 391.}\refname\Cum,\ref{S. Lu, PhD thesis, MIT (1990). A.N. Kirillov,
unpublished.}\refname\Lu,\ref{A.N. Kirillov, P. Mathieu,
 D. S\'en\'echal and M. Walton,``Can fusion coefficients be calculated from
 the depth rule?" LAVAL-PHY-20/92, LETH-PHY-2/92.}\refname\KMSW].  As an
off-shoot of our analysis, we derive  a compact expression for ordinary
$su(3)$ tensor product coefficients.  Our result follows from a new
diagrammatic
method to compute $su(3)$ tensor product coefficients, which is tailor-made for
the analysis of fusion rules.

Let us first recall what fusion rules are.  Consider a Kac-Moody algebra
$\hat{g}$ at level $k$, whose finite Lie algebra is $g$.  Let
$\hat{\lambda}$ be the highest weight of an integrable representation of
$\hat{g}$ at level $k$ and denote its finite part by $\lambda$, which
is then a highest weight of $g$.  $\hat{\lambda}$ is completely specified
by $\lambda$ and $k$.  A fusion rule is the decomposition of a product of two
integrable representations of $\hat{g}_k$ into a sum of integrable
representations. The fusion coefficient $\Nb^{(k)}$ gives the number of times
$C \hat{\nu}$ ( $C$ stands for conjugation) occurs in the product
$\hat{\lambda} \times \hat{\mu}$.  In the limit $k \rightarrow \infty$,
the fusion coefficient $\Nb^{(k)}$ reduces to the ordinary tensor product
coefficient, denoted by $\Nb$ (i.e., $\Nb^{(\infty)} \equiv \Nb$.  Recall
that $\Nb$ gives the number of times the scalar representation occurs in
the triple tensor product $\lambda\otimes\mu\otimes\nu$).  The concept of
fusion rules originates from two dimensional conformal field theory.  In this
 context the formal product $\times$ is related to the radially ordered
operator
product of primary fields [\ref{A.A. Belavin, A.M. Polyakov and A.B.
Zamolodchikov, Nucl. Phys. {\bf B241} (1984) 333.},\ref{E. Verlinde, \np300
(1988) 389.}].  For the WZNW model whose spectrum generating algebra is
$\hat{g}_k$, the primary fields are in one-to-one
 correspondence with integrable representations of
 $\hat{g}_k$ [\ref{V.G.Knizhnik and A. Zamolodchikov, Nucl. Phys.
{\bf B247} (1984) 83.}\refname\KZA,\ref{D. Gepner and E. Witten,
\np278 (1986) 493.}\refname\GW ].

Let us recall the explicit expression for the $\su(2)_k$ fusion coefficients
[\GW,\ref{A.B. Zamolodchikov and V.A. Fateev, Sov. J. Nucl. Phys.
{\bf 43} (1986) 657.}]:

$$\su(2)_{k}:~\Nb^{(k)}=\left\{ \eqalign{ 1 \quad &{\rm if} \quad
|\lambda_{1}-\mu_{1} |\leq\nu_{1}\leq\min\{ \lambda_{1}+\mu_{1},2
k-\lambda_{1}-\mu_{1}\} \cr  &{\rm and} \quad {1 \over 2} (\LLL)\in {\bf Z}_{+}
\cr
 0 \quad & {\rm otherwise}  \cr} \right. \eqlabel\mb$$
Here $\lambda_{1}$ stands for the Dynkin label of $\lambda$ (i.e., $\lambda=
\lambda_{1} \omega^{1}$, where $\omega^{1}$ is the fundamental root of $su(2)$)
and ${\bf Z}_{+}$ denotes the set of non-negative integers.  In this case, it
is
clear that the fusion coefficients are truncated tensor product
coefficients, where the degree of truncation is fixed by the level $k$.  More
precisely, if a coupling $\lambda,\mu,\nu$ is allowed in the finite case (which
means that $|\lambda_{1}-\mu_{1}|\leq\nu_{1}\leq \lambda_{1}+\mu_{1}$ and
${1 \over 2} (\LLL)\in {\bf Z}_{+})$, it also exists in the affine case
as long as $k\geq {1 \over 2} (\LLL).$  There is thus a threshold level
$k_{0}$,
below which the affine coupling is absent and above which it is always present.
  Here clearly $k_0={1 \over 2} (\LLL)$.  For $\su(2)_k$, it is obvious that
the
 fusion coefficients $\Nb^{(k)}$ are completely fixed by the tensor product
coefficients $\Nb$ and $k_{0}$:

$$\su(2)_{k}: \quad\Nb^{(k)} = \left\{
\eqalign{ \Nb & \quad{\rm if}\quad k\geq k_0={1 \over 2} (\LLL) \cr
 0\quad & ~~~{\rm otherwise} \cr}\right. \eqlabel\mc $$

It has been conjectured [\ref{C.J. Cummins, P. Mathieu and
M. Walton, Phys. Lett. {\bf B254} (1991) 386.}\refname\CMW] that for any
Kac-Moody algebra $\hat{g}_{k}$, the fusion coefficients $\Nb^{(k)}$ are
uniquely determined from $\Nb$ and the minimum level $k_0$ at which the various
couplings first appear. \foot{More precisely, in [\CMW] the existence of
$k_{0}$ is conjectured.} Although for $su(2)$ $\Nb$ can only be 0 or 1, in
general it can be greater than one.  Therefore, to the  triplet
$(\lambda,\mu,\nu)$ there correspond $\Nb$ distinct couplings, hence $\Nb$
values of $k_{0}$, one for each distinct coupling. Let us denote these by
$k_{0}^{(i)}, i=1, ..., \Nb$, implementing in this  notation the natural
ordering $k_{0}^{(i)}\leq k_{0}^{(i+1)}$. Then the precise conjectured
relationship between $\Nb^{(k)}$ and the set
$\{\Nb,k_{0}^{(i)}(\lambda,\mu,\nu)
\}$, is \foot{Although at first sight, this result may appear to
be a direct consequence of the Gepner-Witten depth rule [\GW], a closer
analysis demonstrates that this is not the case except for $su(2)$ [\KMSW].}

$$ \Nb^{(k)} = \left\{
\eqalign{\max&(i) \text{such that} k \geq {k_{0}}^{(i)} \text{and}
\Nb\neq 0 \cr    0 & \quad~ \text{if} k<{k_{0}}^{(1)} \text{or} \Nb=0. \cr}
\right. \eqlabel\nbk$$
The conjecture has been proved rigorously for $su(3)$ [\Cum].

We now state our result in the $\su{(3)}_{k}$ case.  At first
we present a closed expression for $\Nb$ which appears to be new.  Given
three $su(3)$ integrable weights $\lambda=(\lambda_{1},\lambda_{2})$
($\lambda_{1,2}$ are the Dynkin labels of $\lambda$), $\mu=(\mu_{1},\mu_{2})$
and $\nu=(\nu_{1},\nu_{2})$, the number of scalars contained in the tensor
product $\lambda\otimes\mu\otimes\nu$ is

$$ \Nb=(k_{0}^{max}-k_{0}^{min}+1)\delta \eqlabel\fus$$
$$\eqalign{&{\rm where} \cr
& k_{0}^{min}=\max\left(\lambda_{1}+\lambda_{2},\mu_{1}+\mu_{2},
\nu_{1}+\nu_{2},{\cal A}-\min(\lambda_{1},\mu_{1},\nu_{1}),
{\cal B}-\min(\lambda_{2},\mu_{2},\nu_{2}) \right) \cr  & k_{0}^{max}=
\min({\cal A},{\cal B}) \cr  & {\cal A}={1 \over 3} [2
(\LLL)+\LLM]=(\lambda+\mu+\nu,\omega^1) \cr & {\cal B}={1 \over 3} [\LLL+2
(\LLM)]=(\lambda+\mu+\nu,\omega^2) \cr   & \delta=\left\{\eqalign{& 1 \text{if}
k_{0}^{max}\geq  k_{0}^{min} \text{and} {\cal A},{\cal B} \in {\bf Z}_{+} \cr
 & 0 \text{otherwise} \cr} \right. \cr} \eqlabel\formule$$ The fusion
coefficients $\Nb^{(k)}$ are given by (\nbk) with the above $\Nb$ and the
following values of $k_{0}^{(i)}$: $$ \{ {k_{0}}^{(i)}\} (\lambda,\mu,\nu)=\{
k_{0}^{min},
 k_{0}^{min}+1,..., k_{0}^{max} \} \eqlabel\niv$$
where $k_{0}^{min}$ and $k_{0}^{max}$ are given in (\formule). That
$$k_{0}^{min} \geq \max (\lambda_{1}+\lambda_{2},\mu_{1}+\mu_{2},
\nu_{1}+\nu_{2}) \eqlabel\slm $$
 simply reflects the fact that the affine extension of the
three weights $\lambda,\mu,\nu$ must be integrable.

These results can be derived directly using the following
diagrammatic approach.  As it will be shown below, $\Nb$ is equal to the
 number of distinct (`bird feet' type) diagrams of the form

$$\trianb{\alpha_1}{\beta_1}{\gamma_1}{p}{\alpha_2}{\beta_2}{\gamma_2}{q}
\eqlabel\dial$$ where $p,q,\alpha_{i},\beta_{i},\gamma_{i}$ are non-negative
integers,
 constrained by the relations

$$\eqalign{& p+\alpha_{1}=\lambda_{1} \quad\quad q+\alpha_{2}=\lambda_{2} \cr
& p+\beta_{1}=\mu_{1} \quad\quad q+\beta_{2}=\mu_{2} \cr
& p+\gamma_{1}=\nu_{1} \quad\quad q+\gamma_{2}=\nu_{2} \cr
\cr}$$

$$ \eqalign{& \alpha_{1}\leq \beta_{2}+\gamma_{2} \quad \beta_{1}\leq
\alpha_{2}+\gamma_{2} \quad \gamma_{1} \leq \alpha_{2}+\beta_{2} \cr
& \alpha_{2} \leq \beta_{1}+\gamma_{1} \quad \beta_{2} \leq \alpha_{1}+
\gamma_{1} \quad \gamma_{2} \leq \alpha_{1}+\beta_{1} \cr} \eqlabel\con$$

$$ \alpha_{1}+\beta_{1}+\gamma_{1}=\alpha_{2}+\beta_{2}+\gamma_{2}$$
(In its complete form, this diagram must appear with three lines joining $p$
to $\alpha_1$, $\beta_1$, $\gamma_1$ and similarly on the other side.)
Distinct
diagrams describe distinct couplings.  To each diagram one associates the
minimum level $k_{0}$

$$k_{0}=p+q+\alpha_{1}+\beta_{1}+\gamma_{1}=p+q+\alpha_{2}+\beta_{2}+\gamma_{2}
 \eqlabel\pq$$
The result (\fus) is a direct consequence of this construction.  On the other
hand the close relation between (\dial) and (\con), the generating function for
tensor product coefficients and that for fusion coefficients, implies (\pq)
directly. Before we turn to an explicit derivation of these results, we give an
example.

 Let us calculate the multiplicity and the corresponding set $\{k_{0}^{(i)}\}$
for the triplet $\lambda=(10,13),~\mu=(11,15), ~\nu=(19,9).$  One finds
$$ \eqalign{& \kmax=\min\{39,38 \}=38 \cr
& \kmin=\max\{23,26,28,38-10,39-9 \}=29 \cr}$$
so that
$$ \Nb=10 \text{and} \{k_{0}^{(i)}\}=\{29,30, ...,38 \}$$ One has for instance
$\Nb^{(32)}$=4. The corresponding diagrams are

        $$\trianb{~9-q}{10-q}{18-q}{~1+q}{13-q}{15-q}{~9-q}{q}$$
$${\rm for}\quad 0 \leq q \leq 9$$
This illustrates clearly the power of formula (\fus) for
computing $su(3)$ tensor product coefficients.

The derivation of (\fus)-(\niv) relies on the generating function for tensor
product coefficients and the Berenstein-Zelevinsky (BZ) triangles.  We
introduce both concepts presently.

The idea of the generating function for tensor product coefficients is based
on the observation that a generic coupling can be decomposed into a finite
number of elementary couplings [\ref{J. Patera and R.T. Sharp,
in {\it Lecture Notes in Physics}, vol. 84, Springer Verlag,
New York 1979.}\refname\PS].  For $su(3)$ there are eight such
elementary couplings:
$$ \eqalign{& E_{1}=(1,0)(0,1)(0,0)~~~ E_{2}=(1,0)(0,0)(0,1)~~~
E_{3}=(0,0)(1,0)(0,1)  \cr
& E_{4}=(0,1)(1,0)(0,0) ~~~E_{5}=(0,1)(0,0)(1,0)~~~E_{6}=(0,0)(0,1)(1,0) \cr
& E_{7}=(1,0)(1,0)(1,0) ~\qquad\qquad\qquad\qquad\qquad\quad
E_{8}=(0,1)(0,1)(0,1)  \cr} \eqlabel\cou$$ (assuming the ordering
$\lambda\mu\nu$).  From a general product of the form

$$E_{1}^{a} E_{2}^{b} E_{3}^{c} E_{4}^{d} E_{5}^{e} E_{6}^{f} E_{7}^{g}
E_{8}^{h} \eqlabel\gen$$
one can read off the Dynkin labels of $\lambda,\mu$ and $\nu$ to be
$$ \eqalign{& \lambda_{1}=a+b+g \qquad \mu_{1}=c+d+g \qquad~\nu_{1}=e+f+g \cr
& \lambda_{2}=d+e+h \qquad \mu_{2}=a+f+h \qquad \nu_{2}=b+c+h \cr}
\eqlabel\vc$$
Notice that to the triplet (1,1)(1,1)(1,1) there corresponds three couplings:
$E_{1} E_{3} E_{5},~ E_{2} E_{4} E_{6}$ and $E_{7} E_{8}$.  But are these all
distinct?  To answer one needs an explicit basis for the couplings.   In
turns out that in any basis, two of these couplings are identical (an
explicit basis is described in the next paragraph, where this result is also
illustrated). Whether $E_{1} E_{3} E_{5}=E_{2} E_{4} E_{6}$, $E_{1} E_{3}
E_{5}=E_{7} E_{8}$ or $E_{2} E_{4} E_{6}=E_{7} E_{8}$ is however dependent upon
an explicit choice of basis.  The invariant result is that there are only two
distinct couplings (of course it is well known that
$N_{(1,1)(1,1)(1,1)}=2$ !).  This illustrates the fact that there are
 redundancies (called syzygies) in the decomposition into elementary
couplings.  Therefore particular products of elementary couplings
must be forbidden.  For instance, in the $su(3)$ case, within the basis for
 which $E_{7} E_{8}=E_{1} E_{3} E_{5}$, one must forbid either $E_{7} E_{8}$ or
$E_{1} E_{3} E_{5}$.  With this proviso, the decomposition of a
general coupling into elementary ones is unique.  Given a complete set of
elementary couplings and a choice of forbidden couplings, one can construct
in a systematic way the generating function for tensor product coefficients
[\ref{C.J. Cummins, M. Couture and R.T. Sharp, J. Phys. A:Math. Gen. {\bf 23}
(1990) 1929.}].
  This is all the information required about generating functions for our
present purpose.

A particularly interesting basis for the couplings is provided by BZ
triangles [\ref{A.D. Berenstein and A.Z. Zelevinsky, Cornell preprint -
technical report 90-60 (1990).}].  They are defined as follows

$$\tri{a_1}{a_2~~}{a_3}{a_4}{a_5}{a_6}{a_7}{a_8}{a_9}$$
where the $a_{i}$'s are non-negative integers fixed by the conditions

$$\matrix{a_1+a_2=\la_1 \quad a_4+a_5=\mu_1 \quad a_7+a_8=\nu_1\cr
 a_3+a_4=\la_2 \quad a_6+a_7=\mu_2 \quad a_9+a_1=\nu_2\cr}$$

$$\matrix{a_2+a_3 = a_6+a_8\cr a_3+a_5 = a_9+a_8\cr a_5+a_6 = a_2+a_9\cr }
\eqlabel\tribz$$
Given a triplet of $su(3)$ integrable weights ($\lambda,\mu,\nu$), the
number of triangles which can be constructed in this way is equal to
$\Nb$.  The triangles associated to $su(3)$ elementary couplings are

$$\matrix{E_1=(1,0)(0,1)(0,0)\cr~\cr\tri010001000}\qquad
\matrix{E_2=(1,0)(0,0)(0,1)\cr~\cr\tri100000000}\qquad
\matrix{E_3=(0,0)(1,0)(0,1)\cr~\cr\tri000010001}$$

$$\matrix{E_4=(0,1)(1,0)(0,0)\cr~\cr\tri000100000}\qquad
\matrix{E_5=(0,1)(0,0)(1,0)\cr~\cr\tri001000010}\qquad
\matrix{E_6=(0,0)(0,1)(1,0)\cr~\cr\tri000000100}$$

$$\matrix{E_7=(1,0)(1,0)(1,0)\cr~\cr\tri010010010}\qquad
\matrix{E_8=(0,1)(0,1)(0,1)\cr~\cr\tri001001001} \eqlabel\zz$$
The triangle associated to a non elementary coupling is simply the sum
of the triangles of the elementary couplings in its decomposition.  In
this way one readily gets
$$\matrix{E_2E_4E_6\cr~\cr\tri100100100}\qquad
\matrix{\neq E_1E_3E_5=E_7E_8\cr~\cr\tri011011011}\eqlabel\EEE$$
Notice that if in the definition of the triangle we had interchanged
the r\^ole of $\mu$ and $\nu$, we would have found instead
$E_{1} E_{3} E_{5} \neq E_{2} E_{4} E_{6}=E_{7} E_{8}$.

Generating function for fusion coefficients have been introduced in [\CMW] (see
also [\ref{L. B\'egin, P. Mathieu and M.A. Walton, J. Phys. A: Math. Gen. {\bf
25} (1992) 135.}]).
  Where it was conjectured that generically there is at least one choice
of  forbidden couplings for which the generating function of fusion
coefficients
 is related in a simple way to that of tensor product coefficients.  In
that case, the value of $k_{0}$ of a coupling which decomposes as
$\prod_{i} E_{i}^{e_{i}}$
reads
$$k_{0}=\sum_{i} e_{i} k_{0} (E_{i}) \eqlabel\nivki$$
where $k_{0} (E_{i})$ is the minimum level at which the elementary coupling
$E_{i}$ first appears.  For $su(3)$, all $k_{0} (E_{i})$ are one.  Furthermore,
in order to reproduce fusion coefficients from (\nivki) it turns out that
$E_{7}
E_{8}$ must $not$ be forbidden (e.g., one must forbid $E_{1} E_{3} E_{5}$ if
the
basis is such that $E_{1} E_{3} E_{5}=E_{7} E_{8}$) [\CMW].  As shown in
[\KMSW]
( see also [\Lu]) the value of $k_{0}$ of a given coupling is actually encoded
in its BZ triangle and reads
$$k_{0}=\max\{a_{1}+\mu_{1}+\mu_{2},a_{4}+\nu_{1}+\nu_{2},a_{7}+
\lambda_{1}+\lambda_{2} \} \eqlabel\nivmax$$

The analysis of fusion rules from BZ triangles is somewhat complicated (cf.
the need to take a maximum over three terms) by the fact that in this basis the
syzygy is not $k_{0}$-homogeneous, i.e., $E_{1} E_{3} E_{5}$ (for which
$\sum k_{0} (E_{i})=3$) is equal to $E_{7} E_{8}$ (with $\sum k_{0}
(E_{i})=2$).   One would expect that a $k_{0}$-homogeneous basis would be
better
adapted to the study of fusion rules.  This motivates us to look for a basis in
which  $E_{1} E_{3} E_{5}=E_{2} E_{4} E_{6}$.

To get a hint on how one should proceed, let us return to (\vc) and see how
Dynkin labels get split off in the case of the BZ triangles.
 The comparison of (\vc) and (\tribz) yields

$$ \matrix{ a_{1}=b \qquad & a_{2}=a+g & a_{3}=e+h & a_{4}=d \qquad & a_{5}=g+c
\cr
 a_{6}=a+h & ~a_{7}=f \qquad  & a_{8}=e+g & a_{9}=h+c & \quad \cr}
\eqlabel\abc$$ from which the hexagon relations ($a_{2}+a_{3}=a_{6}+a_{8}$,
etc)
follow automatically.  Therefore this particular way of breaking Dynkin labels
 in parts ensures that $E_{1} E_{3} E_{5}=E_{7} E_{8}$.  It is clear that
in order to get $E_{1} E_{3} E_{5}=E_{2} E_{4} E_{6}$ different combinations
must be considered.  It is not difficult to see that one actually needs to
combine the integer in (\vc) according to:
$$ \eqalign{& p=g \quad\quad\quad~~ q=h \cr
& \alpha_{1}=a+b \quad \beta_{1}=d+c \quad \gamma_{1}=e+f \cr
& \alpha_{2}=d+e \quad \beta_{2}=a+f \quad \gamma_{2}=b+c  \cr} \eqlabel\fl$$
This leads directly to the diagram (\dial) and the condition specified in
(\con).
  In this basis the elementary couplings are

$$\matrix{E_1=(1,0)(0,1)(0,0)\cr~\cr\tria10000100}\qquad
\matrix{E_2=(1,0)(0,0)(0,1)\cr~\cr\tria10000010}\qquad
\matrix{E_3=(0,0)(1,0)(0,1)\cr~\cr\tria01000010}$$

$$\matrix{E_4=(0,1)(1,0)(0,0)\cr~\cr\tria01001000}\qquad
\matrix{E_5=(0,1)(0,0)(1,0)\cr~\cr\tria00101000}\qquad
\matrix{E_6=(0,0)(0,1)(1,0)\cr~\cr\tria00100100}$$

$$\matrix{E_7=(1,0)(1,0)(1,0)\cr~\cr\tria00010000}\qquad
\matrix{E_8=(0,1)(0,1)(0,1)\cr~\cr\tria00000001} \eqlabel\zlz$$

Now here comes the great simplifying feature of a $k_{0}$-homogeneous basis.
  In that case the $k_{0}$ value of a given coupling can be computed without
caring about which specific couplings have to be forbidden since any choice
gives the same result!  In other words, the $k_{0}$ value of a general product
of the form (\gen) can be computed by simply summing all the exponents
(because $k_{0} (E_{i})$=1 for all $i$) without bothering about the potential
non  uniqueness of the decomposition:

$$k_{0}=a+b+c+d+e+f+g+h \eqlabel\jul$$
Rewriting this result in terms of the data of the diagram (\dial), one gets
directly (\pq).

Now the only thing we have to do is to extract the full set of conditions
ensuring the existence of the diagram (\dial).  From the last condition in
(\con) one has
$$p={\cal A}-{\cal B}+q \eqlabel\jl$$
where $A$ and $B$ are defined in (\fus).
In terms of $p$ or $q$, the six inequalities in (\con) require that
$$p,q \leq\min({\cal A},{\cal B})-\max(\lambda_{1}+\lambda_{2},\mu_{1}+\mu_{2},
\nu_{1}+\nu_{2}) \eqlabel\pqmin$$
Furthermore since each entry in (\dial) must be a positive integer or zero, one
 must have
$$ \eqalign{& p \leq \min(\lambda_{1},\mu_{1},\nu_{1}) \cr
& q \leq\min(\lambda_{2},\mu_{2},\nu_{2}) \cr } \eqlabel\pql$$ (in
addition to the obvious condition that ${\cal A},{\cal B} \in {\bf Z}_{+}$).
Suppose first that ${\cal A} \geq {\cal B}$, so that $q \geq 0$ and $p \geq$
${\cal A}-{\cal B}$.
 Rewriting the first inequality in (\pql) in terms of $q$ and taking into
 account its other two upper bounds, one finds that $$0 \leq q \leq
\min({\cal B}-\max(\lambda_{1}+\lambda_{2},\mu_{1}+\mu_{2},
\nu_{1}+\nu_{2}),\lambda_{2},\mu_{2},\nu_{2},\min(\lambda_{1},\mu_{1},
\nu_{1})-({\cal A}-{\cal B})) \eqlabel\zl$$
In terms of $q$, $k_{0}$ as given in (\pq) becomes
$$k_{0}={\cal B}-q \eqlabel\kaq$$
The number of distinct diagrams (\dial) one can draw (for a fixed triplet
($\lambda,\mu,\nu$)) is thus equal to the number of values that $q$ can
take according to (\zl).  This implies directly
$$\Nb=q^{max}+1 \eqlabel\nombre$$
(Notice that if the minimum on the r.h.s. of (\zl) is negative, there is no
allowed diagram and $\Nb=0$.)  $\Nb$ can clearly be rewritten under the form
$$\Nb=\kmax-\kmin+1 \eqlabel\ww$$
with $\kmax={\cal B}$ (when ${\cal A} \geq {\cal B}$) and $\kmin$ given
in (\formule).
  The requirement of positivity on $q$ can be rephrased as the condition $\kmax
\geq \kmin$.

A similar analysis for ${\cal B} \geq {\cal A}$ (for which $p \geq 0$ and
 $q \geq {\cal B}-{\cal A}$), where now  $$k_{0}={\cal A}-p \eqlabel\www$$
 shows that $\kmax={\cal A}$ while $\kmin$ is the same as before.
The two expressions obtained for $\kmax$ are then equivalent to that given in
(\formule).  Now with either (\kaq) or (\www) one sees directly that to
each distinct coupling  associated to a given triplet $(\lambda,\mu,\nu)$,
there
corresponds a distinct value of $k_{0}^{(i)}$ and that all these values satisfy
$k_{0}^{(i+1)}$=$k_{0}^{(i)}$+1. This completes the proof of (\formule) and
(\niv).

Let us conclude with some comments.  Given a triplet ($\lambda,\mu,\nu$), the
coupling -- described by a diagram of the form (\dial) -- with $largest$
value of $k_0$, is easily found to be that with either $q=0$
(if ${\cal A} \geq {\cal B}$) or $p=0$ (if ${\cal B} \geq {\cal A}$).  All
other
couplings are obtained from it by repeated addition of the ``diagram''
$$\Gamma=\quad\tria{-1}{-1}{-1}{1}{-1}{-1}{-1}{1} \eqlabel\dll$$
In terms of elementary couplings, $\Gamma$ can be written as $E_7 E_8 (E_2 E_4
E_6)^{-1}$ and it contributes $-1$ to $k_0$.  The situation is analogous for
the
BZ triangles.   In that case,
the triangle for the coupling with $smallest$ value of $k_0$ corresponds to
the one for which at least one corner is zero. Which of $a_1,~a_4,~a_7$ is zero
for the triangle of smallest $k_0$, is fully determined by the following
expression
$$\eqalign{&a_1={1 \over 3}{\rm max}(0,\beta,(\beta-\gamma))
\cr &a_4={1 \over 3}{\rm max}(0,\gamma,(\gamma-\beta)) \cr
&a_7={1 \over 3}{\rm max}(0,-\beta,-\gamma) \cr} \eqlabel\retl$$
where
$$\eqalign{ & \beta=3 \lambda_1-3 \mu_2+3 ({\cal B}-{\cal A}) \cr
& \gamma=3 \lambda_2-3 \nu_1+3 ({\cal A}-{\cal B}) \cr} $$
 From this coupling, with smallest possible value of $k_0$, all
others can be obtained by repeated addition of the ``triangle'':
$$\Omega=\tri{~~1}{-1~~}{-1}{1}{-1}{-1}{1}{-1~}{-1~} \eqlabel\wse$$
which decomposes as $E_2 E_4 E_6 (E_7 E_8)^{-1}$ and whose value of $k_0$ is +1
(cf., eq. (\nivmax)).  The number of copies of $\Omega$ which can be
added is limited by the constraint that each entry of the BZ triangle must be a
positive integer or zero.  This gives directly
$$ \Nb={\rm min}(a_2,a_3,a_5,a_6,a_8,a_9)+1 \eqlabel\rrkl $$
where the entries refer to the specific BZ triangle for which
min($a_1,a_4,a_7$)=0 for ($\lambda,\mu,\nu$) fixed.  It is straightforward to
check the full equivalence of this expression with (\fus).  When written at
length, (\rrkl) gives $\Nb-1$ as a minimum over 18 terms which contain each
Dynkin label separately.  Thus for a triplet including one weight with a
vanishing Dynkin label, $\Nb$ is necessarily one, if non-zero.

The result (\fus)-(\niv) could thus have been derived entirely within the BZ
triangle basis.  However the derivation is somewhat simpler in terms of the
diagrams (\dial) and part of the interest of the present derivation lies in the
fact that it displays a new basis for $su(3)$ couplings.

  Let us point out that the product
$N_{\lambda\mu\nu}^{(k)}~N_{{\lambda}^{'}{\mu}^{'}{\nu}^{'}}^{(k+1)}$ gives
directly the fusion rules for the unitary minimal $W_{3}$ models, in which a
primary field is described by a pair of integrable affine weights
$(\hat{\lambda},\hat{\lambda^{'}})$, respectively at level $k$ and $k+1$
[\ref{V.A. Fateev and A.B. Zamolodchikov, Nucl. Phys. {\bf B280} [F S18] (1987)
644.}].

We expect to report elsewhere on the $su(N)$ generalization of this work.
Along that vein, Cummins [\ref{C. Cummins, to appear.  Cummins has also noted
that this result can be seen to follow from Prop. 2.2 of ref. [\GOW].}] has
informed us that using symmetric function techniques it is possible to show
that
$$\kmax \leq {\rm min}[(\lambda+\mu+\nu,\omega^1),
(\lambda+\mu+\nu,\omega^{N-1})] \eqlabel\zxcv $$
for general $\su(N)_k$ fusion rules.  This is
of course consistent with our $su(3)$ expression for $\kmax$ in (\fus).

\vskip12pt
\centerline{\uppercase{acknowledgements}}

We would like to thank C. Cummins, J. Patera and R.T. Sharp for useful
discussions, and  A.N. Kirillov for introducing us to BZ triangles.
Finally we are grateful to D. S\'en\'echal for making available to us his
computer program for fusion rules.


\immediate\closeout\refs \vskip 0.5cm
  \message{References}\input references
\bye